\documentclass[a4paper,twocolumn,aps,pra,showpacs]{revtex4}
\usepackage[latin9]{inputenc}
\usepackage{float}
\usepackage{amsmath}
\usepackage{amssymb}
\usepackage{graphicx}
\usepackage{esint}

\makeatletter

\@ifundefined{textcolor}{}
{%
 \definecolor{BLACK}{gray}{0}
 \definecolor{WHITE}{gray}{1}
 \definecolor{RED}{rgb}{1,0,0}
 \definecolor{GREEN}{rgb}{0,1,0}
 \definecolor{BLUE}{rgb}{0,0,1}
 \definecolor{CYAN}{cmyk}{1,0,0,0}
 \definecolor{MAGENTA}{cmyk}{0,1,0,0}
 \definecolor{YELLOW}{cmyk}{0,0,1,0}
}

\usepackage{mathrsfs}\usepackage{subfigure}\usepackage{multirow}\usepackage[title,titletoc,header]{appendix}

\makeatother
\begin{document}
 \draft
 \title{
 Normal density and moment of inertia of  a moving superfluid }
 \author{Yi-Cai Zhang$^{1}$,  Shu-Wei Song$^2$}
 \author{ Gang Chen$^{3,4,5}$}
\email{chengang971@163.com}
\address{$^{1}$School of Physics and Electronic Engineering, Guangzhou University, Guangzhou
510006, China }
\address{$^2$Wang Da-Heng Collaborative Innovation Center, Heilongjiang Provincial Key Laboratory of
Quantum manipulation and Control, Harbin University of Science and Technology,
Harbin 150080, China}
\address{$^{3}$State Key Laboratory of Quantum Optics and Quantum Optics Devices, Institute of Laser spectroscopy, Shanxi University, Taiyuan 030006, China}

\address{$^4$Collaborative Innovation Center of Extreme Optics,
Shanxi University, Taiyuan, Shanxi 030006, China}
\address{$^5$Collaborative Innovation Center of Light Manipulations and Applications, Shandong Normal University, Jinan 250358, China}
\date{\today}

\begin{abstract}
In this work, the normal density $\rho_n$ and moment of inertia of a moving superfluid are investigated. We find that, even at zero temperature, there exists a finite normal density for the moving superfluid. When the velocity of superfluid reaches sound velocity, the normal density becomes total mass density $\rho$, which indicates that the system losses superfluidity. At the same time, the Landau's critical velocity also becomes zero. The existence of the non-zero normal density is attributed to the coupling between the motion of superflow and density fluctuation in transverse directions. With Josephson relation, the superfluid density $\rho_s$ is also calculated and the identity $\rho_s+\rho_n=\rho$ holds. Further more, we find that the finite normal density also results in a quantized moment of inertia in a moving superfluid trapped by a ring. The normal density and moment of inertia at zero temperature could be verified experimentally by measuring the angular momentum of a moving superfluid in a ring trap.

\end{abstract}

\maketitle
\section{Introduction }
Superfluidity is one of the most striking characteristics of liquid Helium-4 at low temperature, which can flow a narrow tube without any energy dissipation \cite{Kapitza1938,Allen1938}.  The superfluidity usually has close connections with Bose-Einstein condensation \cite{London1938,Bogoliubov1947,Penrose1956}.  Tisza \cite{Tisza1938} and Landau \cite{Landau1941} proposed a two-fluid theory to explain the superfluidity of liquid Helium-4. The basic idea is that the whole liquid with density $\rho$ is divided into two kinds of liquids, i.e., the superfluid part with density $\rho_s$ and the normal part with density $\rho_n$, and the identity $\rho=\rho_s+\rho_n$ holds.  The normal one behaves as usual liquid, which has viscosity and transports entropy; while the superfluid part has no viscosity and can display the superfluidity.

For usual bosonic quantum liquid, for example, liquid Helium-4 or dilute atomic Bose-Einstein condensate, it is believed that, at zero temperature, the superfluid density is the whole liquid density, i.e., $\rho_s=\rho$, and the normal density vanishes, i.e., $\rho_n=0$ \cite{Pines,Leggett}.
If an impurity moves with a velocity, which is  smaller than the Landau's critical velocity, it would not feel any drag force and the motion of the impurity is completely dissipationless.
The superfluidity is usually characterized by a non-zero Landau's critical velocity and a finite superfluid density.
 However, the Landau's critical velocity would decrease if superfluid moves with respect to the laboratory frame.
It is natural to ask the following question:
under this circumstance, does the superfluid density decrease or is there any normal component in the moving superfluid?

In this paper, we try to answer this question by considering a superfluid that moves with a uniform velocity $u$ in a long open-ended tube.
The normal density is calculated with transverse current-current correlation function.   We find that, for a moving superfluid, even at zero temperature, there exists a finite normal density, which is proportional to the square of velocity, i.e., $\rho_n\propto u^2$.  The finite normal density  also brings about a quantized moment of inertia in a ring trap.
The superfluid density is also calculated with an independent method, i.e., Josephson relation. It is found that the superfluid density is proportional to the product of sound velocities of two opposite directions and the identity $\rho_s+\rho_n=\rho$ still holds. The non-zero normal density could be verified by measuring the angular momentum in a ring trap.

 The paper is organized as follows. In Sec.~II, we give the hydrodynamic equations for a moving superfluid (Bose-Einstein condensate) and apply standard quantized rules to get phase and density fluctuation operators in terms of phonon's creation and annihilation operators.  In Sec.~III, the superfluid density is calculated with Josephson's relation, and compare it  with that from phase-twist method. In Sec.~IV, we calculate the normal density with transverse current-current correlation function.  In Sec.~V, a quantized moment of inertia is obtained in a ring trap.  A summary is given in Sec.~VI.

\section{Hydrodynamic  equations}
The Hamiltonian for a dilute Bose atomic gas is \cite{Pethick2002,Pitaevskii}
\begin{align}\label{eqn1}
     \hat{H}&=\hat{H}_0+\hat{V}_{int} ,\notag\\
     \hat{H}_0&=\int d^3\textbf{r}\hat{\psi}^\dag(\textbf{r})\frac{-\hbar^2\nabla^{2}}{2m}\hat{\psi}(\textbf{r}),\notag\\
     \hat{V}_{int}&=\frac{g}{2}\int d^3\textbf{r}\hat{\psi}^\dag(\textbf{r})\hat{\psi}^\dag(\textbf{r})\hat{\psi}(\textbf{r})\hat{\psi}(\textbf{r}),
\end{align}
where $\hat{H}_0$ and $\hat{V}_{int}$ are single-particle Hamiltonian and interaction between atoms, respectively. $\hat{\psi}(\textbf{r})$ is bosonic field operator, $m$ is atomic mass, $g=4\pi\hbar^2 a_s/m$ is interaction strength, and $a_s$ is \emph{s}-wave scattering length.   At zero temperature, superfluidity and Bose-Einstein condensation would occur  and can be characterized by a nonzero order parameter  $\psi\equiv\langle\hat{\psi}(\textbf{r})\rangle=\sqrt{n_0}e^{i\theta}$, with condensate density $n_0$ and phase $\theta$. For weakly interacting Bose gas, the quantum depletion is very small, so the condensate density $n_0$ is approximately equal to the total particle number density $n$, i.e., $n_0\simeq n$ \cite{Pitaevskii}.

The order parameter satisfies the time-dependent Gross-Pitaevskii  equation
\begin{align}\label{eq2}
     i\partial _t \psi(\textbf{r},t)=\frac{-\hbar^2\nabla^2}{2m}\psi(\textbf{r},t)+g|\psi(\textbf{r},t)|^2\psi(\textbf{r},t).
\end{align}
Substituting $\psi(\textbf{r})=\sqrt{n(\textbf{r})}e^{i\theta(\textbf{r})}$ into Eq.~(\ref{eq2}), we obtain the hydrodynamic equations,  which include a continuity equation for mass and a Euler's dynamic equation \cite{Landauhydrodynamic}, i.e.,
\begin{align}\label{qn}
     &\partial_t\rho+ \nabla\cdot \textbf{j}=0,\\
     &\partial_t \textbf{v}+(\textbf{v}\cdot \nabla) \textbf{v}=-\frac{\nabla p}{\rho},\label{qn1}
\end{align}
with mass density $\rho(\textbf{r})=mn(\textbf{r})$, mass current density $\textbf{j}=\rho \textbf{v}$, velocity of superflow $\textbf{v}\equiv \hbar\nabla \theta/m$, and pressure $p=gn^2/2=g\rho^2/(2m^2)$ at zero temperature.  In the derivation of Eqs.~(\ref{qn}) and (\ref{qn1}), we have neglected the quantum pressure term [$\hbar^2\nabla^2 \sqrt{n}/(2m\sqrt{n})$] \cite{Pitaevskii, Tannoudji2011,Pethick2002}. In addition, the velocity is the spatial gradient of phase, so the superfluid velocity must satisfy the condition of non-rotation, i.e.,
   $ \nabla \times \textbf{v}(\textbf{r})=0$.
In the following, we would set the Plank's constant $\hbar=1$, mass $m=1$, and the system volume $V=1$ unless stated otherwise.

 When a superfluid moves along positive \emph{x}-axis direction with velocity $u$ with respect to a  stationary tube, there exists a large energy barrier between the superflow state and the stationary state (true thermodynamic equilibrium state). In such a case,  the moving superfluid becomes a metastable state \cite{Butler1955,Bloch1973,Leggett} (or quasi-equilibrium state \cite {Lee1959}). Near the metastable  state, we linearize the hydrodynamic equations (\ref{qn}) and (\ref{qn1}) as
\begin{align}
     &\partial_t\delta\rho+ u \partial_x\delta\rho+\bar{\rho} \nabla \cdot\delta \textbf{v}=0,\\
     &\partial_t \delta \textbf{v}+u\partial_x \delta \textbf{v}=-\frac{\nabla p}{\bar{\rho}},
\end{align}
with density fluctuation $\delta\rho$,  velocity fluctuation $\delta \textbf{v}$, and average density $\bar{\rho}$. In the following parts of the paper, we will label the average density $\bar{\rho}$ with $\rho$ to simplify the notations.
 Based on the relationship between the velocity and  the phase of condensate, i.e., $\delta \textbf{v}=\nabla\delta\theta$, the linearized hydrodynamic equations become
\begin{align}\label{eqn1}
     &\partial_t\delta\rho+ u \partial_x\delta\rho+\rho \nabla^2 \delta \theta=0,\\
     &\partial_t \nabla\delta\theta+u\partial_x \nabla\delta\theta=-\frac{\partial p}{\rho\partial\rho}\nabla \delta \rho= -\frac{c^2}{\rho}\nabla \delta \rho,\label{eqn2}
\end{align}
where  $c=\sqrt{\partial p/\partial\rho}=\sqrt{gn}$ is the sound velocity when superfluid is at rest.

 From Eqs.~(\ref{eqn1}) and (\ref{eqn2}),  we get the phonon state energy:
 \begin{align}
\omega_\textbf{q}=c_{\textbf{q}}q,
\end{align}
where the sound velocity $c_{\textbf{q}}= c+ u \texttt{cos}\alpha$ and $\alpha$ is the angle between the direction of momentum $\textbf{q}$ and positive $\emph{x}$-axis direction.
For two opposite directions ($\pm \textbf{q}$), we get two different sound velocities,  $c_{\pm \textbf{q}}= c\pm u \texttt{cos}(\alpha)$, and two different energies,  $\omega_{\pm \textbf{q}}= [c\pm u \texttt{cos}(\alpha)]q$.
This is because when the superfluid moves ($u\neq0$), due to the Doppler effects, the sound velocities would be different for two opposite spatial directions \cite{Landauhydrodynamic}.
We see that both the sound velocities and phonon state energies are direction-dependent.
In comparison with the usual stationary case, the Landau's critical velocity of the moving superfluid decreases as $v_{cr}=(\omega_\textbf{q}/q)_{\texttt{min}}=c- u $.

In addition, we can write an effective Hamiltonian for the linearized equations  (\ref{eqn1}) and (\ref{eqn2})
\begin{align}\label{fluidhamiltonian}
&&&H=\frac{1}{2}\int d^3 \textbf{r}\{\rho[(\partial_x \delta \theta)^2+(\partial_y \delta \theta)^2+(\partial_z \delta \theta)^2]\notag\\
&&&+2u\delta\rho \partial_x\delta\theta+\frac{c^2(\delta\rho)^2}{\rho}\}.
\end{align}
In comparison with that of the usual stationary superfluid \cite{zhang2019}, the effective Hamiltonian (\ref{fluidhamiltonian}) has an extra cross term of the density and phase fluctuations,  $2u\delta\rho \partial_x\delta\theta$, which originates from the finite superfluid velocity $u$.
Similarly as the usual stationary superfluid \cite{zhang2019}, by using Poisson brackets $\{\delta\theta(\textbf{r}),\delta \rho(\textbf{r}')\}=-\delta^{3}(\textbf{r}-\textbf{r}')$, we can get the linearized hydrodynamic equations (\ref{eqn1}) and (\ref{eqn2}) from the Hamilton's equation.

After replacing the above classic physical quantities with their corresponding operators, the results of quantization can be obtained by considering the canonical commutator relation $[\delta \hat{\theta}(\textbf{r}),\delta\hat{\rho}(\textbf{r}')]=-i\delta^3(\textbf{r}-\textbf{r}')$ \cite{Lifshitz}.
For example, we can expand the phase operator $\delta\hat{\theta}$ and the density operator $\delta\hat{\rho}$ in terms of  single phonon's annihilation and creation operators, i.e.,
\begin{align}\label{ex}
&\delta\hat{\theta}(\textbf{r},t)\!\!=\!\!\sum_{\textbf{q}\neq0} [A_\textbf{q} C_\textbf{q}e^{i(\textbf{q}\cdot \textbf{r}-\omega_q t)}+A_{\textbf{q}}^{*}C^{\dag}_{\textbf{q}}e^{-i(\textbf{q}\cdot \textbf{r}-\omega_q t)}],\notag\\
&\delta\hat{\rho}(\textbf{r},t)\!\!=\!\!\sum_{\textbf{q}\neq0} [B_\textbf{q} C_\textbf{q}e^{i(\textbf{q}\cdot \textbf{r}-\omega_q t)}+B_{\textbf{q}}^{*}C^{\dag}_{\textbf{q}}e^{-i(\textbf{q}\cdot \textbf{r}-\omega_q t)}],
\end{align}
where $C_\textbf{q}$ $(C^{\dag}_\textbf{q})$ are annihilation (creation) operator for single phonon states, and coefficients  $A_\textbf{q}$ and $B_\textbf{q}$ need to be determined.
From the continuity equation (\ref{eqn1}) (upgrading it as an operator equation), i.e.,
\begin{align}
\label{continue}
\partial _t  \delta\hat{\rho}+u\partial_x\delta\hat{\rho}+\rho\nabla^2\delta\hat{\theta}=0,
\end{align}
 we get
$-icB_\textbf{q}=q\rho A_\textbf{q}$.
From the commutation relation $[\delta \hat{\theta}(\textbf{r}),\delta\hat{\rho}(\textbf{r}')]=-i\delta^3(\textbf{r}-\textbf{r}')$, we get $A_\textbf{q}B^{*}_\textbf{q}=-i/2$ and then  $A_\textbf{q}=-i\sqrt{c/(2\rho q)}$ and $B_\textbf{q}=\sqrt{\rho q/(2c)}$.
Based on Eq.~(\ref{ex}), the density and phase fluctuations in momentum space are written as
\begin{align}\label{rhi}
\hat{\rho}_\textbf{q}\!\!=\!\!\sqrt{\frac{\rho q}{2c}}\left(C_\textbf{q}\!\!+\!C^{\dag}_{-\textbf{q}}\right ),~\hat{\theta}_{\textbf{q}}\!\!=\!\!-i\sqrt{\frac{c}{2\rho q}} \left(C_\textbf{q}\!\!-\!C^{\dag}_{-\textbf{q}}\right).
\end{align}

On the other hand, at low energy, the bosonic field operator can be written as  \cite{Lifshitz}
\begin{align}
\hat{\psi}(\textbf{r})=\langle\hat{\psi}_{}\rangle e^{i\delta\hat{\theta}(\textbf{r})}\simeq \langle\hat{\psi}_{}\rangle \left[1+i\delta\hat{\theta}(\textbf{r})+\cdots\right].
\end{align}
Consequently, in terms of phonon's operators, the field operator in momentum space takes the following form of
\begin{align}
\hat{\psi}_{\textbf{q}}= i\langle\hat{\psi}_{}\rangle \hat{\theta}_{\textbf{q}}=\langle\hat{\psi}_{}\rangle \sqrt{\frac{c}{2\rho q}} \left(C_\textbf{q}-C^{\dag}_{-\textbf{q}}\right).
\label{matrixelement}
\end{align}
The above formulas would be useful in the following discussions.

\section{superfluid density }
 The superfluid density can be given by the Josephson relation \cite{Josephson1966,Holzmann,Bogoliubov,zhang2018}, i.e.,
 \begin{align}\label{Jos}
 \rho_s(\textbf{q})=-\underset{q\rightarrow0}{\texttt{lim}}\frac{n_0}{q^2 G(\textbf{q},0)},
\end{align}
where $G(\textbf{q},\omega)=\sum_n[\frac{|\langle0|\hat{\psi}_{\textbf{q}}|n\rangle|^2}{\omega-\omega_{n0}}-\frac{|\langle0|\hat{\psi}^{\dag}_{\textbf{q}}|n\rangle|^2}{\omega+\omega_{n0}}]$ is normal Green's function with excitation state $|n\rangle$ and excitation  energy  $\omega_{n0}=E_n-E_0$ \cite{Lifshitz}, and $n_0=|\langle\hat{\psi}\rangle|^2$  is condensate density.

With Eq.~(\ref{matrixelement}), we calculate the matrix element of $\hat{\psi}_{\textbf{q}}$ ($\hat{\psi}^{\dagger}_{\textbf{q}}$) between the ground state $|0\rangle$ and the single phonon state $|n\rangle=|\textbf{q}\rangle=C^{\dag}_{\textbf{q}}|0\rangle$, i.e., $\langle0|\hat{\psi}_{\textbf{q}}|\textbf{q}\rangle=\langle\hat{\psi}_{}\rangle \sqrt{\frac{c}{2\rho q}} $, $\langle0|\hat{\psi}^{\dag}_{\textbf{q}}|\textbf{q}\rangle=-\langle\hat{\psi}_{}\rangle^* \sqrt{\frac{c}{2\rho q}}$,  and thus the Green's function $G(\textbf{q},0)=-[\frac{|\langle\hat{\psi}\rangle|^2c}{2\rho c_{+\textbf{q}} q^2}+\frac{|\langle\hat{\psi}_{}\rangle|^2c}{2\rho  c_{-\textbf{q}} q^2}]$. Therefore, the superfluid density
\begin{align}
\rho_s(\textbf{q})=\frac{\rho c_{+\textbf{q}}c_{-\textbf{q}}}{c^2}=\rho[1-\frac{u^2\texttt{cos}^2(\alpha)}{c^2}].
\label{rhos}
\end{align}
In the above equation, we have used the fact that the single phonon states have dominant contributions to the Green's function $G(\textbf{q},0)$, while the contributions of the multiple phonon states can be neglected  as $q\rightarrow0$.
 For a usual stationary superfluid, i.e., $u=0$ ($c_{+\textbf{q}}=c_{-\textbf{q}}=c$), the superfluid density $\rho_s(\textbf{q})$ is equal to the total density $\rho$. However, for a moving superfluid ($u\neq0$), the superfluid density is smaller than the total density. When the velocity of superfluid is equal to the sound velocity, i.e., $u=c$, the superfluid density of \emph{x}-axis
 direction would vanish, i.e., $\rho_s(\hat{\textbf{x}})=0$ [$\alpha=0$ in Eq.~(\ref{rhos})], where $\hat{\textbf{x}}$ is a unit vector in positive \emph{x}-axis direction.
In the next section, with current-current correlation function, we will show that when superfluid density becomes zero, the normal density reaches its maximum value, i.e., $\rho_n=\rho$.

Due to the anisotropy of the sound velocity, the superfluid density is usually a  second-order tensor in three dimensional space \cite{zhang2018}, namely, $\rho_{s}=\textbf{diag}\{\rho_{s}(\hat{\textbf{x}}),\rho_{s\perp}=\rho,\rho_{s\perp}=\rho\}$. The superfluid density in \textbf{q}'s direction can be given by a  tensor contraction, i.e.,
\begin{align}
\rho_{s}(\textbf{q})=\hat{\textbf{q}}\cdot\rho_s\cdot \hat{\textbf{q}}=\rho_{s}(\hat{\textbf{x}})\texttt{cos}^{2}(\alpha)+\rho_{s\perp}\texttt{sin}^{2}(\alpha),
\end{align}
where $\hat{\textbf{q}}=\textbf{q}/q$ is the unit vector in $\textbf{q}$-direction.

In many cases, the superfluid density is usually defined with a phase twist method \cite{Fisher}, i.e.,
 \begin{align}
\tilde{\rho}_{s}(\hat{\textbf{x}})=\frac{2\Delta E}{(\delta \theta/L)^2},
\end{align}
where $\delta \theta$ is phase difference between two ends of tube filled with liquid, $L$ is tube length, and
$\Delta E$ is energy cost due to the phase gradient between two ends.
We further assume that the density fluctuation is negligible  ($\delta\rho\equiv0$), and the phase gradient is constant ($\partial_x \delta\theta=\delta\theta /L$) for a superflow state. Thus, the energy in Eq.~(\ref{fluidhamiltonian}) is
\begin{align}
\Delta E=\frac{\rho  (\delta \theta/L)^2}{2}.
\end{align}
So, we obtain the superfluid density
$\tilde{\rho}_s(\hat{\textbf{x}})=\rho$.
The above result shows the superfluid density defined by using the phase twist method is the total density $\rho$, which is not consistent with the result from the Josephson relation. Only when $u=0$ ($c_{+\textbf{q}}= c_{-\textbf{q}}$), these two methods yield the same result.

\section{normal density  }
  In this section, we will calculate the normal density $\rho_n$ with current-current correlation function \cite{Pines}.
  Let's assume there exists a long straight open ended tube filled with a moving superfluid with a uniform speed $u$ along positive \emph{x}-axis direction. In addition, we also assume the tube moves at a small velocity $v$ along \emph{x}-axis direction.   If the superfluid reaches equilibrium (or quasi-equilibrium) with the tube wall, the normal part would be dragged by the tube wall and moves with tube at the same velocity $v$. Then, the resulting mass current density due to the motion of the tube would be
  \begin{align}
\delta j_x=\rho_n v.
\end{align}
The above equation can be viewed as a definition of the normal density $\rho_n$.
Here the normal density $\rho_n$ can be calculated with the transverse current-current correlation function \cite{Baym,normaldensity},
 \begin{align}
  \rho_n(\hat{\textbf{x}})\!\!=\!\!\underset{q\rightarrow0}{\texttt{lim}}\sum_n\!\!\left[\!\frac{|\langle0|\hat{j}_{x,\textbf{q}=\textbf{q}_\bot}|n\rangle|^2}{\omega_{n0}}\!\!+\!\!\frac{|\langle0|\hat{j}_{x,-\textbf{q}_\bot}|n\rangle|^2}{\omega_{n0}}\!\right],
\end{align}
where $\hat{j}_{x,\textbf{q}}$ is current fluctuation operator in momentum space and vector $\textbf{q}_\perp$ is ``transverse" with respect to \emph{x}-direction, i.e.,  $ \hat{\textbf{x}}\cdot  \textbf{q}_\perp =0$ ($\textbf{q}_{\perp}$ and $\hat{\textbf{x}}$ are perpendicular to each other ).  The calculation of the normal density with the transverse current-current correlation function amounts to ask whether the superfluid system could have response to a transverse probe or not \cite{Pines}.

The current fluctuation operator $\hat{j}_{x,\textbf{q}}$ can be obtained conveniently from hydrodynamic equations. For example, from Eq.~(\ref{continue}),
 we read off the effective current fluctuation operators for low energy phonon states:
\begin{align}
\delta\hat{j}_x=u\delta\hat{\rho}+\rho\partial_x\delta\hat{\theta},~\delta\hat{j}_y=\rho\partial_y\delta\hat{\theta},~\delta\hat{j}_z=\rho\partial_z\delta\hat{\theta}.
\end{align}
In momentum space, they take the form of
\begin{align}
\hat{j}_{x,\textbf{q}}\!=\!u\hat{\rho}_\textbf{q}+i\rho q_x \hat{\theta}_\textbf{q},~\hat{j}_{y,\textbf{q}}\!=\!i\rho q_y\hat{\theta}_\textbf{q},~\hat{j}_{z,\textbf{q}}\!=\!i\rho q_z\hat{\theta}_\textbf{q}.
\end{align}
In contrast to the usual stationary superfluid, here we see the current fluctuation operator $\hat{j}_{x,\textbf{q}}$ includes an extra density fluctuation term $u\hat{\rho}_\textbf{q}$,  which would result in a finite normal density $\rho_n$.
Without loss of generality, taking the transverse direction $\textbf{q}=\textbf{q}_{\perp}=q\hat{\textbf{y}}$ ($\hat{\textbf{y}}$ is a unit vector of \emph{y}-axis direction) for instance, we get the normal density of \emph{x}-axis direction,
\begin{align}\label{normal}
\rho_n(\hat{\textbf{x}})=\rho\frac{u^2}{c^2}=\rho^2 u^2\kappa,
\end{align}
with the compressibility $\kappa$, which satisfies a sum rule of density-density correlation function  \cite{Pines1966}
\begin{align}\label{compress}
\rho^2\kappa&=\underset{q\rightarrow0}{\texttt{lim}}\sum_n\left [\frac{|\langle0|\rho_{\textbf{q}=\textbf{q}_\bot}|n\rangle|^2}{\omega_{n0}}+\frac{|\langle0|\rho_{-\textbf{q}_\bot}|n\rangle|^2}{\omega_{n0}}\right]\notag\\
&=\underset{q\rightarrow0}{\texttt{lim}}\frac{2|\langle0|\rho_{\textbf{q}=\textbf{q}_\bot}|\textbf{q}_{\perp}\rangle|^2}{\omega(\textbf{q}_\bot)}=\frac{\rho}{c^2}.
\end{align}
In the compressibility sum rule, we have used Eq.~(\ref{rhi}) and the fact that the single phonon state dominates the contribution as $q\rightarrow0$ and the contributions of phonon states of two transverse directions ($-\textbf{q}_{\perp}$ and $+\textbf{q}_{\perp}$) are the same.
Taking Eqs.~(\ref{rhos}) and (\ref{compress}) into account, the superfluid density can be written, in terms of two sound velocities and compressibility, as
\begin{align}
\rho_s(\textbf{q})=\frac{\rho c_{+\textbf{q}}c_{-\textbf{q}}}{c^2}=\rho^2c_{+\textbf{q}}c_{-\textbf{q}}\kappa,
\end{align}
which is consistent with Eq.~(15) in Ref.~\cite{normaldensity}

The above results show that, for a moving superfluid ($u\neq0 $), there would exist a finite normal density even at zero temperature. The existence of the finite normal density has a close connection with density fluctuation.  More precisely speaking, the finite normal density results from the coexistence of motion of superflow and density fluctuation in transverse directions.
It is worthwhile to emphasis that the above conclusion is quite universal for a generic superfluid system.
This is because Eq.~(\ref{normal}) indicates that, as long as  the moving superfluid has a finite positive compressibility, i.e., $\kappa>0$ (the system is thermodynamically stable against density fluctuation), there would exist a finite normal density.
In addition,
from Eqs.~(\ref{rhos}) and (\ref{normal}),
 we see the identity $\rho_s(\hat{\textbf{x}})+\rho_n(\hat{\textbf{x}})=\rho$ still holds (in \emph{x}-direction).

The existence of the finite normal density can be understood as follows. Since the density fluctuation $u\delta\rho$ appears in the current fluctuation $\delta j_x$, the occurrence of the density fluctuation always causes a current change. For example, when a mass density fluctuation occurs in a superfluid, such a  density fluctuation would drift away along main stream, which results in the change of current.
When $u=c$, the normal density is equal to the total density, i.e., $\rho_n=\rho$, which shows that the total density becomes normal,  the superfluid density vanishes [for the case of  $\alpha=0$ in Eq.~(\ref{rhos})],  and the system losses its superfluidity.
At the same time, the vanishing of the superfluid density is also consistent with the vanishing of the Landau's critical velocity, i.e., $v_{cr}=c-u=0$.

  In  the above discussions, we have assumed the Landau's critical velocity is determined solely by the sound velocity (e.g., for Bose-Einstein condensate of dilute atomic gas). In addition, a moving impurity with a velocity $u>c$ in a stationary condensate would begin to dissipate energy and experience a drag force \cite{Pitaevskii12004}.
However, if the Landau's critical velocity is not determined by the sound velocity (e.g., by roton minimum in liquid helium), when the superfluid moves faster than the Landau's critical velocity, there may exist condensate of roton excitations \cite{Pitaevskii1980} or occur a second order phase transition between a uniform phase to a spatially periodic state \cite{Pitaevskii1984,Baym2012}.

\section{Quantized  momentum of inertia}
In the following part, we recover the notations: mass $m$, Plank's constant $\hbar$ and system volume $V$.
Here let's assume the superfluid moves in a ring trap with width $d$, and the radius of ring is $R$ \cite{Leggett}. At the same time, the trap geometrical parameters satisfy $d<<R$.  Due to the constraint of ring geometry, the allowed circulation of the superfluid velocity must take  quantized value $n 2\pi \hbar/m$ \cite{Pitaevskii}, and the velocity should be given by
\begin{align}\label{quantized}
u=\frac{n \hbar}{m R},
\end{align}
where integer $n=0,1,2,3,\cdots$.

Assuming the angular velocity of trap perpendicular to the plane of ring is $\delta\omega$ and the rotating axis is through the center of ring, therefore the linear velocity $v=\delta\omega R$. Due to existence of the finite normal density, the normal part would be dragged by the rotating trap.
 The angular momentum arising from the slow rotation is
\begin{align}
\delta L=\delta\omega I,
\end{align}
with moment of inertia
\begin{align}
I=\rho_n V R^2.
\end{align}
Combining it with Eqs.~(\ref{normal}) and (\ref{quantized}),  the moment of inertia turns out to be
\begin{align}
I=\frac{n^2M\hbar^2}{m^2c^2}=2n^2M\xi^2,
\end{align}
where $M=\rho V$ is total mass of superfluid and $\xi=\hbar/(\sqrt{2}mc)$ is healing length \cite{Pitaevskii}. The above equation shows the moment of inertia is quantized in unit of $M\xi^2$, which is the direct consequence of quantized circulation of velocity field in a non-simply connected region.

For a stationary superfluid ($u=0$ and $n=0$), the normal density vanishes, the  moment of inertia and angular momentum would be zero, which corresponds to the Hess-Fairbank effect \cite{Leggett} that the superfluid part always keeps stationary in liquid Helium as long as the angular velocity of rotating bucket is small enough \cite{Hess}. However,
when $u\neq0$, a moving superfluid can respond to the slow rotation and has a
non-zero moment of inertia even at zero
temperature.

To avoid the formation of vortices in the ring, the velocity of superfluid $u$ should be smaller than the Feynman's critical velocity,
$v_{F}=\hbar ln(d/\xi)/(md)$ \cite{Feynman}.
On the other hand, the superfluid velocity should  be also smaller than the sound velocity, i.e., $u<c$.
So the integer $n$ in Eq.~(\ref{quantized}) should satisfy
$n<R ln(d/\xi)/d$ and $n<mcR/\hbar$.
The long lifetime persistent current states (a moving superfluid) in ring geometry have been experimentally created \cite{Ryu2007,Ramanathan2011,Moulder2012,Eckel2014}. In principle, the finite normal density and non-zero  moment of inertia  could be detected by imparting non-zero angular momentum into atom gas \cite{Cooper2010}.

\section{summary}
In conclusion, we find that there exists non-zero normal density for a moving superfluid even at zero temperature,  which is proportional to the square of velocity of the superfluid. The finite normal density can be attributed to the coupling between the motion of superflow and density fluctuation in transverse direction.
 The finite normal density  also results in a quantized moment of inertia in a moving superfluid trapped by a ring.
In addition, We find that the superfluid density is proportional to product of sound velocities of two opposite directions and  the identity  $\rho_s+\rho_n=\rho$ still  holds. It is expected that the finite normal density and non-zero moment of inertia of a moving superfluid in ring trap may be measured experimentally by using optical method in atomic gases in near future.

\section*{Acknowledgements}

Yi-Cai Zhang thanks Shizhong Zhang for useful discussions. This work was
supported by the NSFC under Grants No.~11874127 and No.~11674200.
 Yi-Cai Zhang also
acknowledges the support of a startup grant from Guangzhou University. S. W. Song was supported by Heilongjiang Provincial Natural Science Foundation of
China with No. LH2019A015, the Fundamental Research Fundation for Universities
of Heilongjiang Province with No. LGYC2018JC005.

\appendix*


\begin{thebibliography}{10}

\bibitem{Kapitza1938} P. Kapitza, Viscosity of liquid helium below the $\lambda$-point, Nature (London) \textbf{141}, 74 (1938).
 \bibitem{Allen1938} J. F. Allen and A. D. Misener, Flow of liquid helium II, Nature (London) \textbf{141}, 75 (1938).
  \bibitem{London1938} F. London, The $\lambda$-phenomenon of liquid helium and the Bose-Einstein degeneracy, Nature (London) \textbf{141}, 643 (1938).




\bibitem{Bogoliubov1947} N. N. Bogoliubov, On the theory of superfluidity, J. Phys. (USSR) \textbf{11}, 23 (1947).
  \bibitem{Penrose1956} O. Penrose and L. Onsager, Bose-Einstein condensation and liquid helium, Phys. Rev. \textbf{104}, 576 (1956).
\bibitem{Tisza1938} L. Tisza, Transport phenomena in Helium II, Nature (London) \textbf{141}, 913 (1938).
  \bibitem{Landau1941} L. D. Landau, The theory of superfuidity of Helium II, J. Phys. U.S.S.R. \textbf{5}, 71 (1941).

\bibitem{Pines} D. Pines, P. Nozi\`{e}res, The theory of Quantum Liquids, Vol. 2 (Addison-Wesley, Redwood City, CA, 1990).
\bibitem{Leggett} A. J. Leggett, Quantum Liquids (Oxford University Press, 2006).

\bibitem{Pethick2002}C. J. Pethick and H. Smith, Bose-Einstein Condensation in dilute gases  (Cambridge University Press, 2002).
\bibitem{Pitaevskii} L. Pitaevskii and S. Stringari, Bose-Einstein Condensation  (Oxford University Press, 2003).
\bibitem{Tannoudji2011} C. Cohen-Tannoudji and D. Gu\'{e}ry-Odelin, Advances in Atomic Physics: An Overview,  Chap. 22 (World Scientific, 2011).


  \bibitem{Butler1955} S. T. Butler and J. M. Blatt,  Nonequilibrium Nature of the Superfluid State,  Phys. Rev. \textbf{100}, 495 (1955).


  \bibitem{Bloch1973} F. Bloch,  Superfluidity in a ring,  Phys. Rev. A \textbf{7}, 2187 (1973).

 \bibitem{Lee1959} T. D. Lee and C. N. Yang, Low-Temperature Behavior of a Dilute Bose System of Hard Spheres. II. Nonequilibrium Properties,  Phys. Rev. \textbf{113}, 1406 (1959).

\bibitem{Landauhydrodynamic}  L. D. Landau and E. M. Lifshitz, Course of theoretical Physics, Vol. 6, Fluid Mechanics  (Pergamon Press, 1987).


\bibitem{zhang2019} Y.C. Zhang, C.F. Liu, B. Xu, G. Chen, W. M. Liu , Two-fluid theory for a superfluid system with anisotropic effective masses, Phys. Rev. A \textbf{99}, 043622 (2019).




 \bibitem{Lifshitz} E. M. Lifshitz and L. P. Pitaevskii, Course of theoretical Physics, Vol. 9, Statistical Physics part 2 (Pergamon Press, 1980).

\bibitem{Josephson1966} B. D. Josephson,  Relation between the superfluid density and order parameter for superfluid He near $T_c$,  Phys. Lett. \textbf{21}, 608 (1966).
\bibitem{Bogoliubov} N. N. Bogoliubov, Lectures on Quantum Statistics (Gordon and Breach, New York, 1970).
\bibitem{Holzmann} M.Holzmannand, G.Baym, Helicity Modulus, Condensate superfluidity and infrared structure of the single-particle Green's function:
The Josephson relation, Phys. Rev. B \textbf{76}, 092502 (2007).

\bibitem{zhang2018} Y.C. Zhang, Generalized Josephson relation for conserved charges in multicomponent bosons, Phys. Rev. A \textbf{98}, 033611 (2018).

\bibitem{Fisher} M. E. Fisher, M. N. Barber, D. Jasnow, Helicity Modulus, Superfluidity, and Scaling in Isotropic Systems, Phys. Rev. A \textbf{8}, 1111 (1973).

\bibitem{Baym} G. Baym, in Mathematical Methods in Solid State and Superfuid Theory, edited by R. C.
Clark and E. H. Derrick (Oliver and Boyd, Edinburgh, 1967).


\bibitem{normaldensity} Y.C. Zhang, Z.Q. Yu, T. K. Ng, S. Zhang, L. Pitaevskii, S. Stringari, Superfluid density of a spin-orbit-coupled Bose gas, Phys. Rev. A \textbf{94}, 033635 (2016).
We note that there should be a factor $\rho^2$ rather than $\rho$ in right-hand side of Eq.~ (15) in Ref. \cite{normaldensity}.



  \bibitem{Pines1966} D. Pines  and  P. Nozi\`{e}res, Theory of Quantum Liquids, Vol. I (Benjamin,
New York, 1966).




\bibitem{Pitaevskii12004} G. E. Astrakharchik, L. P. Pitaevskii, Motion of a heavy impurity through a Bose-Einstein condensate,
 Phys. Rev. A \textbf{70}, 013608 (2004).






\bibitem{Pitaevskii1980} S. V. lordanskii, L. P. Pitaevskii, Bose condensation of moving rotons, Sov. Phys. Usp. \textbf{23}, 317-8 (1980).

\bibitem{Pitaevskii1984} L. P. Pitaevskii,  Layered structure of superfluid $^4$He with supercritical motion, JETP Lett. \textbf{39}, 511-14 (1984).

\bibitem{Baym2012} G. Baym, C. J. Pethick, Landau critical velocity in weakly interacting Bose gases,
 Phys. Rev. A \textbf{86}, 023602 (2012).



 \bibitem{Hess} G. B. Hess, W. M. Fairbank, Measurements of angular momentum in superfluid Helium, Phys. Rev. \textbf{19}, 216 (1967).
\bibitem{Feynman} R. Feynman, Statistical Mechanics, A set of Lecture (Benjamin, 1972).


\bibitem{Ryu2007} C. Ryu, M. F. Andersen, P. Clad\'{e}, V. Natarajan, K. Helmerson, W. D. Phillips, Observation of Persistent Flow of a Bose-Einstein Condensate in a Toroidal Trap,
 Phys. Rev. Lett. \textbf{99}, 260401 (2007).

\bibitem{Ramanathan2011}A. Ramanathan, K. C. Wright, S. R. Muniz, M. Zelan, W. T. Hill III, C. J. Lobb, K. Helmerson, W. D. Phillips, G. K. Campbell, Superflow in a toroidal Bose-Einstein condensate: an atom circuit with a tunable weak link,
 Phys. Rev. Lett. \textbf{106}, 130401 (2011).




\bibitem{Moulder2012} S. Moulder, S. Beattie, R. P. Smith, N. Tammuz, Z. Hadzibabic, Quantised supercurrent decay in an annular Bose-Einstein condensate,
Phys. Rev. A \textbf{86}, 013629 (2012).



\bibitem{Eckel2014} S. Eckel, J. G. Lee, F. Jendrzejewski, N. Murray, C. W. Clark, C.
 J. Lobb, W. D. Phillips, M. Edwards, G. K. Campbell, Hysteresis in a quantized superfluid `atomtronic' circuit, Nature (London) \textbf{506}, 200 (2014).




\bibitem{Cooper2010}
N. R. Cooper, Z. Hadzibabic, Measuring the superfluid fraction of an ultracold atomic gas, Phys. Rev. Lett. \textbf{104}, 030401 (2010).
\end{thebibliography}
\end{document}